\documentclass[aps,prl,twocolumn,letterpaper,longbibliography]{revtex4-1}
\usepackage{graphicx}
\usepackage{bm}
\usepackage{amssymb}
\usepackage{color}
\usepackage{amsmath}
\usepackage{ulem}
\usepackage{bbold}
\usepackage[mathlines]{lineno}
\usepackage[colorlinks=True,citecolor=blue,linkcolor=blue,allcolors=blue]{hyperref}
\usepackage{enumitem}
\usepackage[utf8]{inputenc}

\def\blue{\textcolor{blue}}
\def\red{\textcolor{red}}
\begin{document}

\title{Floquet higher-order Weyl and nexus semimetals}
\author{Weiwei Zhu}
\thanks{These authors contribute equally.}
\affiliation{Department of Physics, National University of Singapore, Singapore 117542, Singapore.}

\author{Muhammad Umer}
\thanks{These authors contribute equally.}
\affiliation{Department of Physics, National University of Singapore, Singapore 117542, Singapore.}

\author{Jiangbin Gong}
\email{phygj@nus.edu.sg}
\affiliation{Department of Physics, National University of Singapore, Singapore 117542, Singapore.}

\begin{abstract}
This work reports the general design and characterization of two exotic, anomalous nonequilibrium topological phases.
In equilibrium systems, the Weyl nodes or the crossing points of nodal lines
may become the transition points  between
higher-order and first-order topological phases defined on two-dimensional slices, thus featuring both hinge Fermi arc and surface Fermi arc.   We  advance this concept by presenting a strategy to obtain, using time-sequenced normal insulator phases only,
  Floquet higher-order Weyl semimetals and Floquet higher-order nexus semimetals, where the concerned topological singularities in the three-dimensional Brillouin zone border anomalous two-dimensional higher-order Floquet phases.
  The fascinating topological phases we obtain are previously unknown and can be experimentally studied using, for example, a three-dimensional lattice of coupled ring resonators.
\end{abstract}

\maketitle
{\it Introduction.---}Owing to their rich features not available in equilibrium counterparts, the great potential of nonequilibrium topological phases is being widely recognized~\cite{Kitagawa2010,Lindner2011, Ho2012,Rudner2013,Rechtsman2013,Gomez-Leon2013,Kundu2013,Lababidi2014,FoaTorres2014,Hubener2017,Rudner2020,Wintersperger2020,McIver2020}.    Given that even an otherwise topologically trivial system can be converted into a topological nontrivial phase via periodic driving~\cite{Gong2008,Kitagawa2010}, nonequilibrium strategies that utilize the time dimension significantly expand the domain of topological phases and at the same time lead us to many possible applications
of topological matter in photonics~\cite{Rechtsman2013,Hu2015,Gao2016,Leykam2016,Maczewsky2017,Mukherjee2017}, acoustics~\cite{Peng2016}, quantum information etc~\cite{RadityaGong2018,Bomantara2020}.
Three-dimensional (3D) periodically driven (Floquet) topological phases include
Floquet Weyl semimetals~\cite{Wang2014,Wang2016,Zhou2016,Zhang2016a,Yan2016,Chan2016,Zhang2016,Bomantara2016,Hubener2017,Yan2017,Ezawa2017,Bucciantini2017,Liu2017,Li2018,Chen2018,Sun2018,Higashikawa2019,Zhu2020a,Umer2020} and Floquet topological insulators~\cite{Ladovrechis2019,He2019,Schuster2019}.
Floquet Weyl semimetals may support a single Weyl point, thus constituting an excellent platform to investigate the chiral magnetic effect~\cite{Sun2018,Higashikawa2019}. The so-called Floquet Hopf insulator,
whose full topological description requires one $\mathbb{Z}$ type Hopf linking number and another intrinsically dynamic $\mathbb{Z}_2$ invariant~\cite{He2019,Schuster2019}, is another fascinating example that has advanced the notion of topological insulators.

Our focus here is on Floquet higher-order topological phases.  Floquet higher-order topological insulators
(FHOTIs) have already been designed \cite{Bomantara2019,Rodriguez-Vega2019,Seshadri2019,Nag2019,Peng2019a,Plekhanov2019,Ghosh2020b,Hu2020,Huang2020,Peng2019,Chaudhary2020, Bomantara2020a,Bomantara2020,Zhu2021,Zhu2020,Ghosh2020a,Zhang2020,Ghosh2020,Liu2020,Nag2020,Zhu2021a,Vu2021}.
Inspired by these progresses, we report two exotic Floquet topological semimetal phases with boundary-of-boundary states, namely, Floquet higher-order Weyl semimetals (HOWSs) and Floquet  higher-order nexus semimetals (HONSs).  Our results are of general interest because (i) the obtained HOWS and HONS phases are not describable by standard equilibrium approaches (such as quadrupole moments \cite{Benalcazar2017,Schindler2018}) when describing the concerned higher-order topological phases, (ii) they can be generated by a simple periodically driven bipartite lattice whose instantaneous Hamiltonian is {\it always} a normal insulator;  and (iii) our  general flexible design does not rely on precise tuning of the system parameters.  The first feature clearly distinguishes this work from two early studies on Floquet HOWSs of which the results under high-frequency driving do have equilibrium analogs \cite{Ghorashi2020,Rui2020}. The second and third features indicate experimental feasibility to confirm our theoretical predictions.

\begin{figure}
\includegraphics[width=\linewidth]{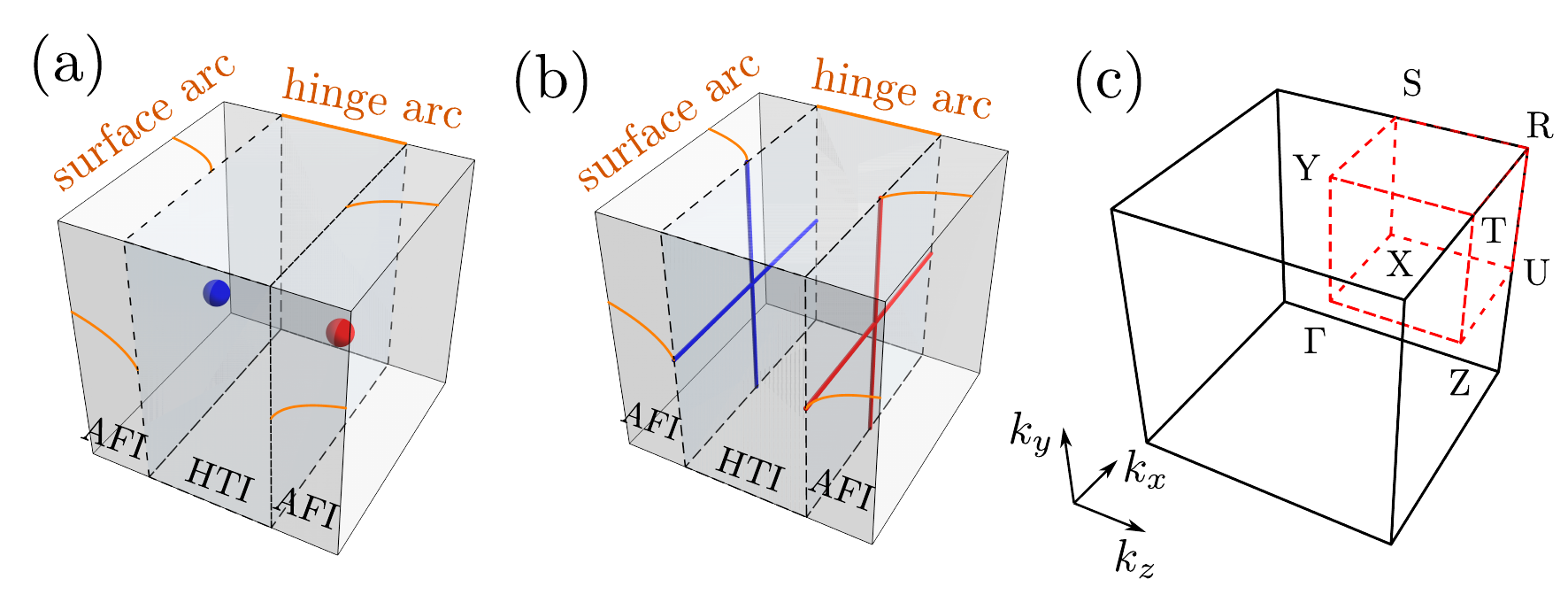}
\caption{ (a) HOWSs with surface Fermi arcs and hinge Fermi arcs (orange lines). The Weyl point with charge $+1$ ($-1$) is marked by blue (red) ball. (b) Same as in (a) but for HONSs. The winding number surrounding the nodal line marked blue (red) tube is $+1$ ($-1$).  (c) The high symmetry points in the Brillouin zone are marked by $\Gamma$, $\mathrm{X}$, $\mathrm{Y}$, $\mathrm{S}$, $\mathrm{Z}$, $\mathrm{U}$, $\mathrm{T}$ and $\mathrm{R}$.}
\label{schematic}
\end{figure}

We first introduce the concept of HOWS and HONS. Due to the Nielsen-Ninomiya theorem \cite{Nielsen1981a,Nielsen1981}, the minimal Weyl semimetal in a lattice accommodate two Weyl nodes of opposite charges, as shown in Fig.~\ref{schematic}(a). In a traditional Weyl semimetal, the Weyl node represents the phase transition point between Chern insulator and normal insulator of some 2D slices, so it supports the well-known surface Fermi arc upon opening up its boundary along one direction. By contrast, in a HOWS \cite{Ghorashi2020,Wang2020,Wei2020} the Weyl node is the phase transition point between a Chern insulator and a higher-order topological insulator (HOTI). As such, in addition to the expected surface Fermi arc,  the system possesses the hinge Fermi arc when opening up its boundaries along two directions.   On the other hand, HONS phases are featured by two crossed nodal lines, with the associated phase transition lines splitting 2D slices into Chern insulators and HOTI phases, so that the system also supports surface Fermi arcs and hinge Fermi arcs.   The physics is much more involving in nonequilibrium situations, because the Weyl nodes in Floquet HOWS and the crossed nodal lines in Floquet HONS feature exotic phase transitions between hybrid topological insulators (HTI), anomalous Floquet topological insulators (AFI), and anomalous Floquet higher-order topological insulator (AFHOTI),  as summarized in Figs.~\ref{schematic}(a) and ~\ref{schematic}(b). The said anomalous topological phases here have zero Chern numbers and vanishing multipolarization, and therefore only topological invariants capturing characteristics of dynamical evolution within one complete period may fully describe them \cite{Rudner2013,Zhu2021,Yu2021}.

\begin{figure}
\includegraphics[width=\linewidth]{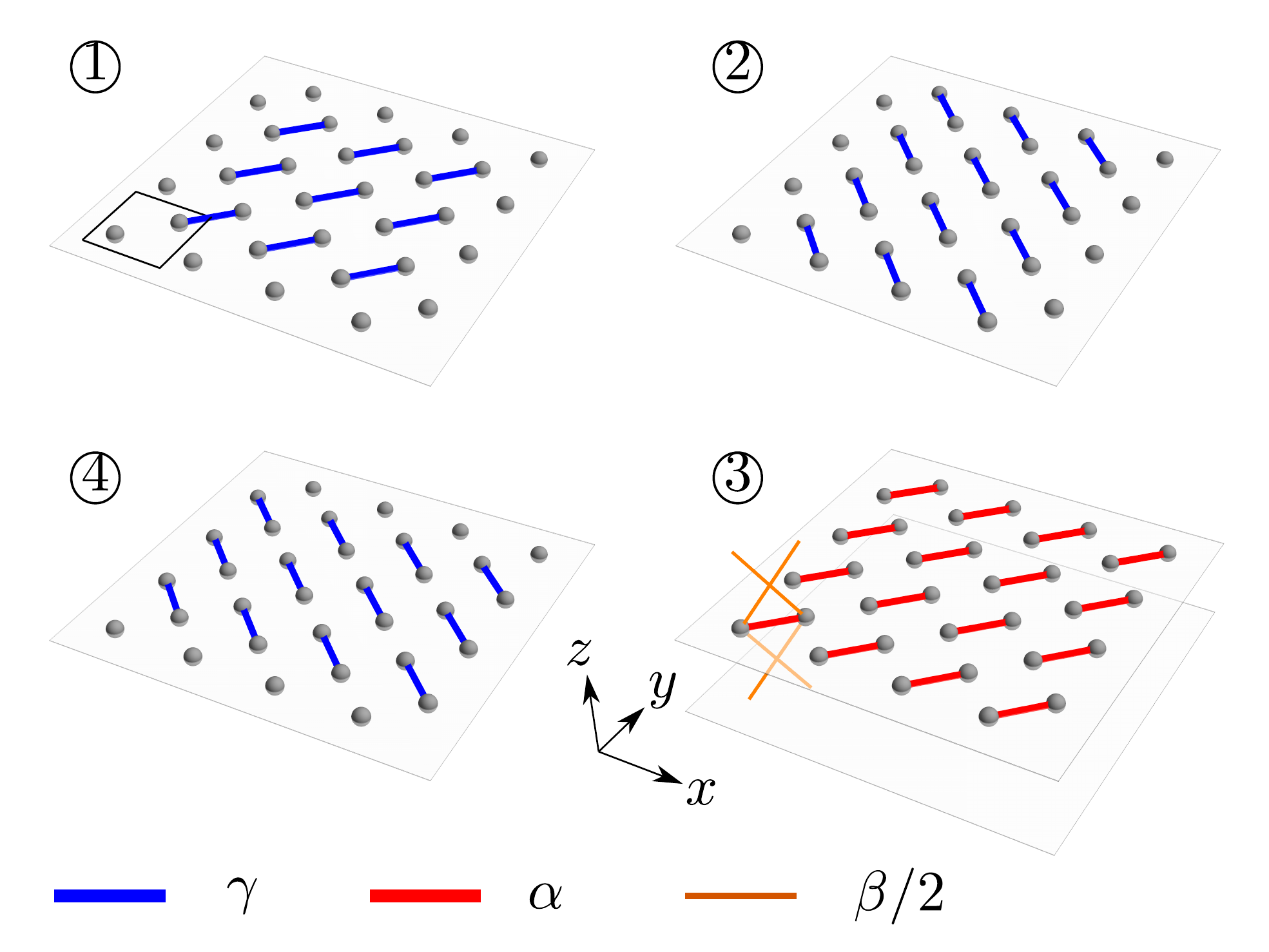}
\caption{A periodically driven bipartite lattice subject to four modulation steps. Step $\textcircled{1}$, $\textcircled{2}$ and $\textcircled{4}$ allows for intercellular coupling of strength $\gamma$ within the $xy$ layer. Step $\textcircled{3}$  introduces intracellular coupling of strength $\alpha$ and interlayer coupling of strength $\beta/2$.  Only the interlayer coupling of a primitive unit cell is shown.  The  rectangle in step $\textcircled{1}$ indicates one unit cell.}
\label{model}
\end{figure}

{\it Model.---} Consider a periodically driven bipartite lattice shown in Fig.~\ref{model}. This minimal model is composed of four time steps and only the nearest-neighbor coupling is required. Each unit cell consists of two sub-lattice degrees of freedom. 
In step $\textcircled{1}$, $\textcircled{2}$ and $\textcircled{4}$, the lattice is fully dimerised and only turns on inter-cellular coupling of strength $\gamma$ within the $xy$ layer. This kind of two-dimensional setup is equivalent to an experimentally realized platform of two-dimensional coupled ring resonators \cite{Zhu2021}. In step $\textcircled{3}$, the system allows for intracellular coupling of strength $\alpha$ and inter-layer coupling of strength $\beta/2$.  In each time step, the instantaneous Hamiltonian is only a normal insulator under the condition $\gamma\neq0$ and $|\alpha|>|\beta|$.  The time-dependent Bloch Hamiltonian is then given by,
\begin{eqnarray}
 H(\mathbf{k},t)=\left\{
\begin{array}{cc}
H_{1}(\mathbf{k})&0<t\leq T/4\;,\\
H_2(\mathbf{k})&T/4<t\leq T/2\;,\\
H_3(\mathbf{k})&T/2<t\leq 3T/4\;,\\
H_4(\mathbf{k})&3T/4<t\leq T\;,
\end{array}
\right.
\label{eq1}
\end{eqnarray}
where,
\begin{equation}
H_{m}(\mathbf{k})=\gamma(e^{ib_m\cdot\mathbf{k}}\sigma^{+}+h.c.)\;,
\label{eq2}
\end{equation}
for $m=1,2,4$. $\sigma^{\pm}=(\sigma_{x}\pm i\sigma_{y})/2$, where $\sigma_{x,y,z}$ are Pauli matrices; and the vectors $\mathbf{b}_{m}$ are given by $\mathbf{b}_{1}=(a,a,0)$, $\mathbf{b}_{2}=(a,0,0)$ and $\mathbf{b}_{4}=(0,a,0)$, where $a$ is the lattice constant. Moreover,
\begin{equation}\label{eq3}
 H_{3}(\mathbf{k})=\theta(k_z)\sigma_x\;,
\end{equation}
where $\theta(k_z) = \alpha + \beta\cos(b_3\cdot\mathbf{k})$ with $\mathbf{b}_{3}=(0,0,a)$.
The expression $\theta(k_z)$ here suggests that it already fully captures the effect of interlayer coupling along $z$ direction and so becomes important to digest our design below.
The Floquet states can be found from,
\begin{equation}\label{eq4}
  U_T(\mathbf{k},t_0)\, |\phi(\mathbf{k},t_0)\rangle = e^{-i\varepsilon(\mathbf{k}) T} |\phi(\mathbf{k},t_0)\rangle\;,
\end{equation}
where $U_T(\mathbf{k},t_0) \equiv \mathcal{T}\mathrm{exp}\big[-i\int_{t_0}^{t_0+T} H(\mathbf{k},\tau)\,d\tau\big]$ is the one-period time evolution operator, $\mathcal{T}$ is the time-ordering operator, and $t_0$ is a reference time.

A Weyl semimetal phase requires the breaking of time reversal or inversion symmetry \cite{Wan2011}. The time reversal symmetry in our driven model is clearly broken.  As to other aspects of the symmetry, our system obeys particle-hole $C^{-1}H(\mathbf{k},t)C=-H^{*}(-\mathbf{k},t)$ and inversion $\mathcal{I}^{-1} H(\mathbf{k},t) \mathcal{I} = H(-\mathbf{k},t)$ symmetry, where $C = \sigma_{z}$ and $\mathcal{I} = \sigma_{x}$. The particle-hole symmetry promises that the Weyl points appear at quasienergy zero or $\pi/T$ while the inversion symmetry ensures that the Weyl points of opposite charge possess the same quasienergy.  Furthermore, the system obeys two fold rotation symmetry about $z$ axis, i.e., $C_{2z}^{-1} H(\mathbf{k},t) C_{2z} = H(\mathbf{\tilde{k}},t)$ with $C_{2z}=\sigma_x$ and $\mathbf{\tilde{k}}=(-k_x,-k_y,k_z)$. This symmetry guarantees that there is always such two-fold rotation symmetry even when we reduce the system to 2D slices at an arbitrary fixed value of $k_z$. To explain our design,  we note that the resultant 2D slices can support HTI (where $\pi/T$ gap supports chiral edge mode and zero gap supports topological corner mode), AFI (where both $\pi/T$ gap and zero gap support chiral edge modes), and AFHOTI (where both  $\pi/T$ gap and zero gap support topological corner modes)~\cite{Zhu2021,Zhu2020}.
As analyzed below, these intriguing properties point towards genuinely anomalous HOWS phases.

\begin{figure}
\includegraphics[width=\linewidth]{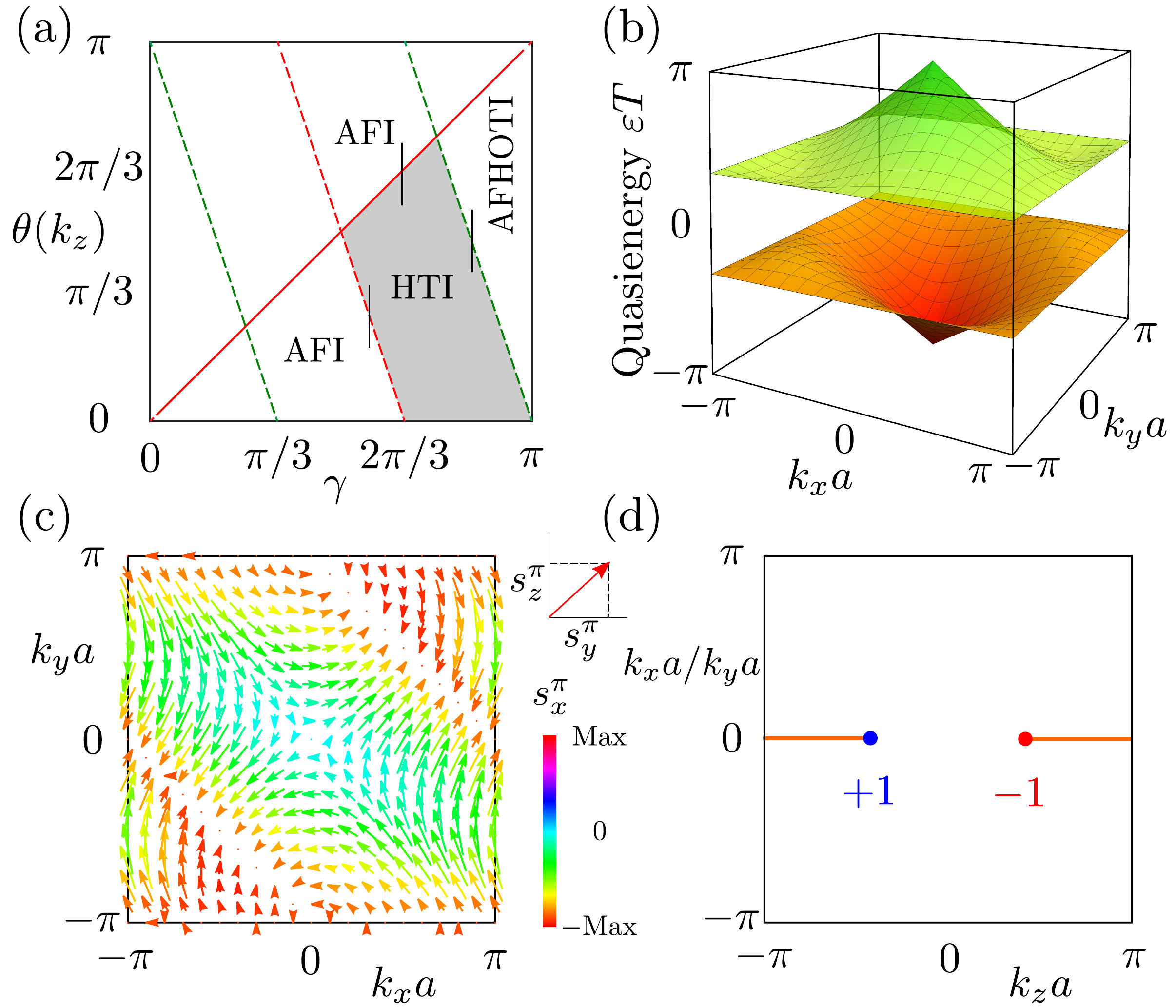}
\caption{ (a) Phase diagram for system's $2$D slices at fixed $k_z$. (b) Quasienergy spectrum and (c) pseudo-spin texture in $k_x-k_y$ plane at $k_z^c$.
Color represents the value of $x$-component and the direction indicates the other two components. (d) Surface Fermi arc of Floquet HOWS with quasienergy $\pi/T$ in $k_x-k_z$ plane or $k_y-k_z$ plane. The parameters for (b)(c)(d) are $\alpha=9\pi/20$, $\beta=\pi/5$ and $\gamma=5\pi/6$. The HTI phase is marked by gray area in (a). In (a), dashed line (solid) is the phase transition point at $\Gamma$ ($k_x=\pm\pi/a$ and $k_y=\pm\pi/a$). Green (red) line means topological phase transition at quasienergy zero ($\pm\pi/T$).
}
\label{Weyl}
\end{figure}

{\it Floquet HOWSs and SFA.---} The phase diagram for a $2$D slice at a fixed $k_z$ is shown in Fig.~\ref{Weyl}(a), in terms of possible values $\theta(k_z)$ and $\gamma$. Topological phase transition happens at $\Gamma$ point ($k_x=0~\wedge~k_y=0$) with quasienergy zero marked by red dashed line, at $\Gamma$ point but with quasienergy $\pi/T$ marked by green dashed lines or at two high symmetry lines $k_x=\pi/a~\vee~k_y=\pi/a$ with quasienergy zero marked by red solid line (see Supplementary Materials). The HTI phase is marked by gray area in Fig.~\ref{Weyl}(a), enclosed by three boundaries: the green dashed line is the phase boundary between HTI and AFHOTI; the red (dashed and solid) lines represent the phase boundary between HTI and AFI.

To appreciate the flexibility of our design, one may now look into what happens if we tune $\alpha$ and $\beta$ values and hence change the possible range of $\theta(k_z)$.  As $k_z$ varies from $-\pi/a$ to $\pi/a$, one can easily let $\theta(k_z)$ go across the phase boundaries  between the above mentioned phases of the 2D slices [see black lines in Fig.~\ref{Weyl}(a) as three examples at three different values of $\gamma$]. Precisely at the phase boundary, the bulk band of the original 3D system closes itself.   For example, consider $\alpha=9\pi/20$, $\beta=\pi/5$ and $\gamma=5\pi/6$ such that $\pi/4\leq \theta(k_z)\leq 13\pi/20$. This range of $\theta(k_z)$
covers the phase transition point at $\theta=\pi/2$ (determined by $\gamma$).  Inverting the expression $\theta(k_z)$ one finds that the bulk band closes at  $k_{z} = \pm k_z^c = \pm{\arccos(1/4)/a}$, with quasienergy $\pi/T$, yielding two band closing points $(0,0,\pm k_z^c)$ as possible Floquet Weyl nodes.

Such possible locations of Weyl nodes can also be understood by symmetry eigenvalues at high symmetry momentum points (HSMPs). At $k_z=0$, there are four HSMPs, $\Gamma$, $\mathrm{X}$, $\mathrm{Y}$ and $\mathrm{S}$ shown in Fig.~\ref{schematic}(c). The inversion symmetry eigenvalues at those points are $(+1,+1,+1,+1)$. At $k_z=\pi/a$, there are also four HSMPs, $\mathrm{Z}$, $\mathrm{U}$, $\mathrm{T}$ and $\mathrm{R}$ shown in Fig.~\ref{schematic}(c). The corresponding symmetry eigenvalues are $(-1,+1,+1,+1)$.  When the value of $k_z$ change from 0 to $\pi/a$, the symmetry eigenvalues must change along the $\Gamma-\mathrm{Z}$ direction. Furthermore, the system's two-fold rotation symmetry promises the band inversion to occur along $\Gamma-\mathrm{Z}$, which again indicates that Weyl nodes are at $k_x=0$ and $k_y=0$.

To finally confirm the emergence of Weyl nodes, we present the linear quasienergy dispersion at the band crossing points $(0, 0, \pm{k_{z}^{c}})$ in Fig.~\ref{Weyl}(b). The topology of the band crossing points can also be investigated by the pseudo-spin texture,
\begin{equation}\label{eq5}
  s_i^j = \left[j/T-\varepsilon(\mathbf{k})\right]\langle\phi^\varepsilon(\mathbf{k},0)|\sigma_i|\phi^\varepsilon(\mathbf{k},0)\rangle\;,
\end{equation}
where $\sigma_i$ is the Pauli matrix with $i=x,y,z$; $j=0$ ($j=\pi$) for the band crossing points with quasienergy zero ($\pi/T$). Choosing the band below quasienergy $j/T$, the pseudo-spin texture thus defined at $k_{z} = k_{z}^{c}$ is depicted in Fig.~\ref{Weyl}(c). There vanishing pseudo-spin components are observed at $\Gamma$ point. It is also seen that $s_x^\pi$ is zero [in Fig.~\ref{Weyl}(c), blue indicates zero $s_x^\pi$] in the vicinity of $\Gamma$ point.
(See Supplementary Materials for more analysis). Circling around $\Gamma$ point, the winding number of the nonzero spin textures $(s_y^\pi,s_z^\pi)$ is $-1$.  Using these results, one then finds that
the chirality of the identified Weyl nodes to be
 $ q=-\mathrm{sgn}\left[\left.\frac{\partial\theta(k_z)}{\partial k_z}\right|_{k_z=\pm k_z^c}\right]$.
The charge is $+1$ for Weyl node at $(0,0,-k_z^c)$ and $-1$ for $(0,0,+k_z^c)$ (See Supplementary Materials).  Connecting the two Weyl points with opposite charges, we also obtain dispersion-free surface Fermi arc states with quasienergy $\pi/T$, in the $k_x-k_z$ plane with $y$ direction open or in the $k_y-k_z$ plane with $x$ direction open. They are represented as straight lines in Fig.~\ref{Weyl}(d).

\begin{figure}
\includegraphics[width=\linewidth]{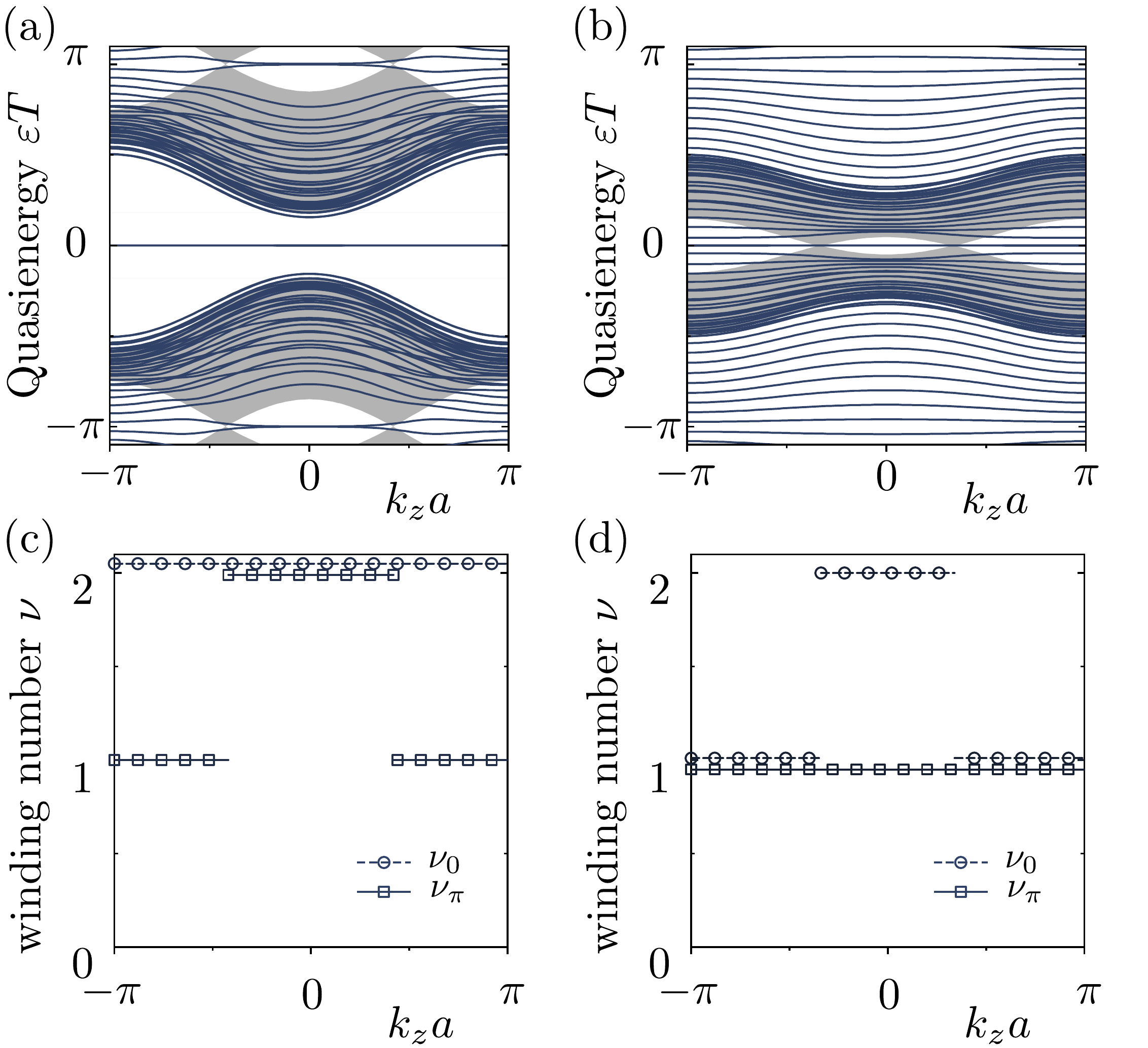}
\caption{(a)(b) Spectrum of a finite structure with open boundary condition along $x,y$ and periodic boundary condition along $z$. The Weyl points are located at quasienergy $\pi/T$ in (a) and at quasienergy zero in (b). (c)(d) shows the corresponding winding numbers defined in the main text. The parameters for (a)(c) are $\alpha=9\pi/20$, $\beta=\pi/5$ and $\gamma=5\pi/6$. The parameters for (b)(d) are $\alpha=\pi/5$, $\beta=\pi/10$ and $\gamma=7\pi/12$. The bulk band is marked by gray.}
\label{spectrum}
\end{figure}

{\it Higher-order topology and hinge Fermi arc.---} One peculiarity of the obtained Floquet Weyl semimetal phases is their hinge Fermi arc states when taking open boundary condition along both $x$ and $y$ directions. The quasienergy spectrum under OBC (PBC) in $x-y~(z)$-direction, with the same parameters as in Fig.~\ref{Weyl}(d), is shown in Fig.~\ref{spectrum}(a). In the $\pi/T$ quasienergy gap, the system supports transport surface modes for $|k_z|>k_z^c$, but supports dispersionless hinge modes for $|k_z|<k_z^c$, thus confirming the nature of HOWSs.  The zero gap features a higher-order topological insulator phase that is not of our interest here.  Remarkably, these exotic topological features cannot be described by traditional bulk topological invariants. Indeed,
unlike equilibrium systems, a Floquet system contains two quasienergy gaps (zero and $\pi/T$), possibly with markedly different behaviours, thus requiring additional topological indices for complete characterization.  Because the HOWSs obtained here are designed from slices of 2D Floquet topological phases with their anomalous phase diagram shown in Fig.~\ref{Weyl}(a),  topological features of our HOWSs associated with each quasienergy gap may be further digested  using an earlier approach already well applied to such 2D systems \cite{Zhu2021,Zhu2020}. That is, for a fixed $k_z$,  we look into a gap winding number by further reducing the dimension along the direction $k_x=k_y=k$, which gives effectively one dimensional system that has chiral symmetry~\cite{Zhu2021,Zhu2020}.  One can then choose $t_0=T/8$ as the starting reference time to define a time-symmetric frame, with the associated Floquet operator $U_T({\bf k}, t_0)$ denoted as $U_T(k,T/8)$.
Further rewriting $U_T(k,T/8)$ as
\begin{equation}\label{eq7}
  U_T(k,T/8)=\sigma_zF^{\dag}\sigma_zF\;,
\end{equation}
with $F$ being a $2\times 2$ matrix, we quickly obtain the winding numbers associated with two quasienergy gaps~\cite{Asboth2014} (as a function of $k_z$):
\begin{eqnarray}
  \nu_0 &=& \int_{-\pi/a}^{\pi/a} \partial(\arg F_{12})\;,\\
  \nu_\pi &=& \int_{-\pi/a}^{\pi/a} \partial(\arg F_{22})\;.
\end{eqnarray}
The winding numbers $(\nu_0, \nu_\pi)$ defined above take integer values $\in \{0,1,2\}$ in our system.  A zero winding number indicates that the concerned gap is topological trivial.  On the other hand,  winding number being 1 and 2 respectively indicates that the existence of one chiral edge mode and a pair of counter-propagating edge modes at each edge (see Supplementary Materials).  In the presence of counter-propagating edge modes the system's feature is determined by the actual boundary condition along $x$ and $y$ directions. If the boundary condition is chosen as of Fig.~\ref{model}, the counter-propagating edge modes are gapped and the sliced 2D system supports corner modes and hence the original 3D system supports hinge modes (see Supplementary Materials). In Fig.~\ref{spectrum}(c), we show the obtained winding numbers $(\nu_0, \nu_\pi)$ for the parameter values corresponding to Fig.~\ref{spectrum}(a). It can be observed that when the scanned $k_z$ values cross that of the above-identified Weyl nodes, the winding number $\nu_\pi$ makes a jump according to the node chirality.  This not only offers a practical means to characterize the system's topology with dimension reduction, but also confirms again our HOWS design.  The constant winding number $\nu_0=2$ shown in Fig.~\ref{spectrum}(c) is also consistent with the shown corner modes in the 0 quasienergy gap in Fig.~\ref{spectrum}(a).

It is also interesting to look into another representative case with parameter values $\alpha=\pi/5$, $\beta=\pi/10$ and $\gamma=7\pi/12$ such that the Weyl nodes of opposite chirality appear at zero quasienergy for $(k_{x}, k_{y}, k_{z}) = (0, 0, \pm k^c_z)$ with $k_z^c=\arccos(1/2)/a$. The quasienergy spectrum under OBC (PBC) in $x-y~(z)$ spatial directions is shown in Fig.~\ref{spectrum}(b) and hinge Fermi arc is now observed between two Weyl nodes at zero quasienergy. The HOWSs is then characterized by jumps in the winding number $\nu_0$ from $1$ to $2$ as shown in Fig.~\ref{spectrum}(d), which fully captures the feature of hinge Fermi arc and the chirality of two Weyl nodes.

{\it Floquet higher-order nexus semimetal.---}  The general design strategy discussed above is versatile because it can equally generate higher-order nexus semimetals (HONSs), featured by the crossing point of two nodal lines \cite{Heikkila2015,Chang2017,Tang2020}.  HONS phases are somewhat analogous to HOWS phases
because they also possess hinge Fermi arc in addition to surface Fermi arc.  To realize Floquet HONS, one can now fully exploit the red solid line in Fig.~\ref{Weyl}(a) as the phase boundary between HTI and AFI.
Unlike other phase boundaries, in this special case the phase boundary always lies at two highly symmetric lines ($k_x=\pi/a$ and $k_y=\pi/a$) with zero quasienergy.  Just as how we tune $\alpha$ and $\beta$  values in the case of HOWS, here we do the same, but targeting at the outcome that as $k_z$ changes, the resultant $\theta(k_z)$ values cross the red solid line in Fig.~\ref{Weyl}(a).  When this is indeed the case, then the 3D system represents a Floquet nexus semimetal phase as the band crossing always occurs along $k_x=\pi/a$ and $k_y=\pi/a$.

\begin{figure}
\includegraphics[width=\linewidth]{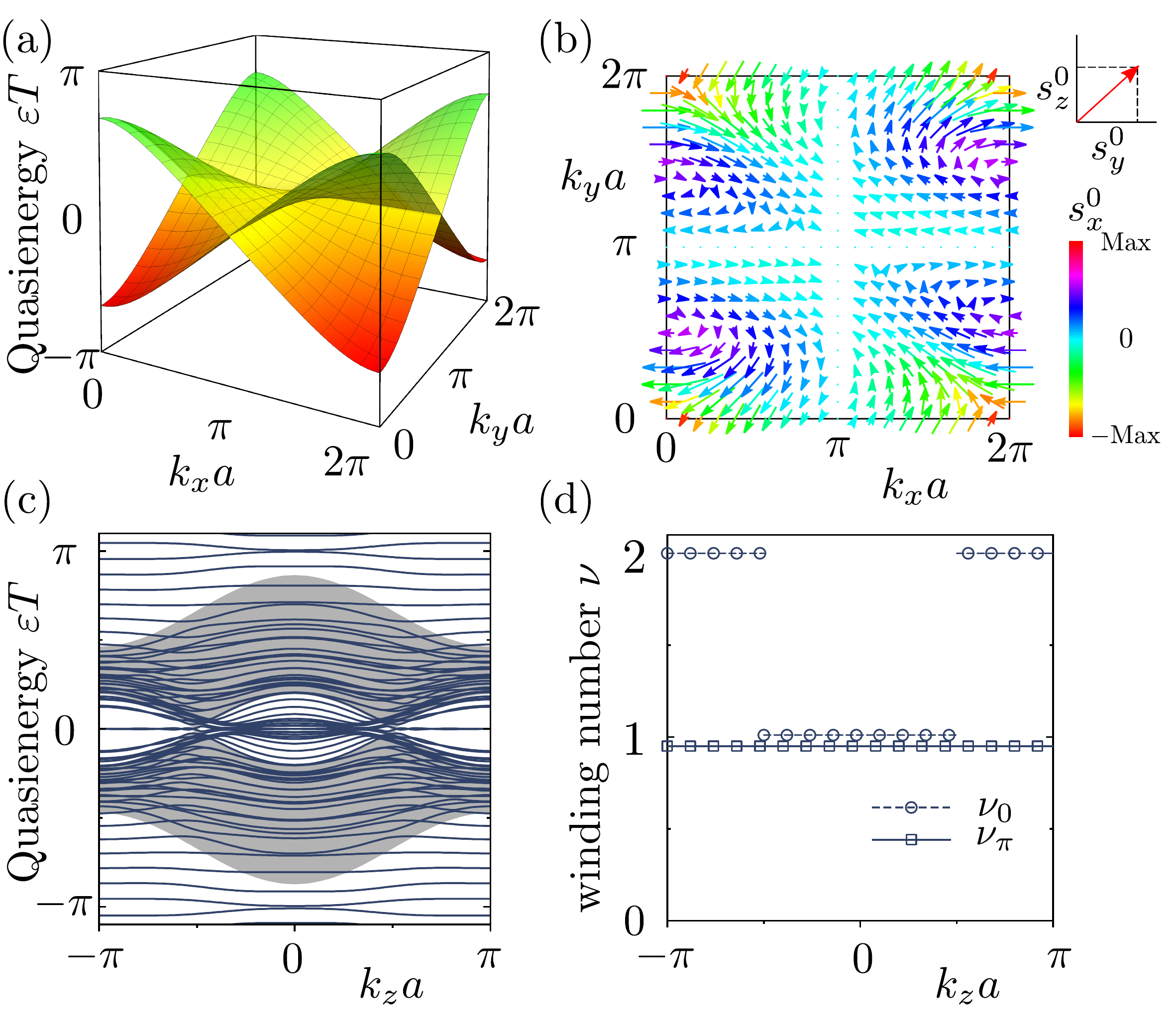}
\caption{Floquet HONS, its spectrum and the behavior of topological invariants. (a) Quasienergy band at $k_z^c$. Two nodal lines are located at $k_x=\pi/a$ and $k_y=\pi/a$, yielding the nexus at $\mathbf{k}=(\pi/a,\pi/a,k_z^c)$. (b) pseudo-spin texture for nexus fermions in (a). (c) Spectrum of a finite structure as a function of $k_z$. (d) The winding numbers for two band gaps as a function of $k_z$. The parameters are $\alpha=2\pi/3$, $\beta=\pi/5$ and $\gamma=2\pi/3$. The bulk band is marked by gray.}
\label{nexus}
\end{figure}

As an example, consider $\alpha=2\pi/3$, $\beta=\pi/5$ and $\gamma=2\pi/3$ such that the nodal lines appear at zero quasienergy at $k_z=\pm k_z^c$, with  $k_z^c=\pi/(2a)$. These nodal lines are shown in Fig.~\ref{nexus}(a) where the nexus (crossing point of the two nodal lines) can be observed at $k_{x} = k_{y} = \pi/a$. A symmetry analysis is in order. With $k_z=0$, the symmetry eigenvalues of $\sigma_x$ at $\Gamma$, $\mathrm{X}$, $\mathrm{Y}$ and $\mathrm{S}$ are $(-1,-1,-1,-1)$. With $k_z=\pi/a$, the symmetry eigenvalues at $\mathrm{Z}$, $\mathrm{U}$, $\mathrm{T}$ and $\mathrm{R}$ are $(-1,+1,+1,+1)$. Hence, as  $k_z$ changes from 0 to $\pi/a$, the band inversion occurs along $\mathrm{X}-\mathrm{U}$, $\mathrm{Y}-\mathrm{T}$ and $\mathrm{S}-\mathrm{R}$, which are indeed special points on the nodal lines. Moreover, the pseudo-spin texture is shown in Fig.~\ref{nexus}(b). It is observed that all the components of pseudo-spin are zero at the two highly symmetric lines $k_x=\pi/a$ and $k_y=\pi/a$.  In order to characterize the nodal lines $k_x=\pi/a$  and $k_y=\pi/a$, we use a Berry phase based winding number.   Specifically, for $k_x=\pi/a+r\cos(\xi)$ ($k_y=\pi/a+r\cos(\xi)$), $k_y$ ($k_x$) is fixed and $k_z=\pm k_z^c+r \sin(\xi)$, the following winding number
\begin{equation}\label{eq10}
  W=-\frac{i}{2\pi}\int_0^{2\pi}d\xi\langle\phi(\xi)|\partial_\xi|\phi(\xi)\rangle\;.
\end{equation}
around the closed loop is quantized.  This winding number is given as $\pm{1}$ for the nodal lines at $\mp k_z^c$ (see Supplementary Materials).

By our design, the crossing point of two nodal lines (nexus) now have implications for higher-order topological edge states. Indeed, the spectrum in Fig.~\ref{nexus}(c) clearly indicates that the nexus represents the phase transition point from HTI to AFI for 2D slices.  Therefore, upon opening the system's boundary,  at quasienergy zero the system changes from supporting hinge Fermi arc to supporting surface Fermi arc.  To confirm the higher-order topology, we examine again the above-defined gap winding number $\nu_0$ in Fig.~\ref{nexus}(d).  There it is seen that $\nu_0$
 changes from $1$ to $2$ when $|k_z|$ values cross that of a nexus.

{\it Conclusion and discussion.---} This work has provided a rather general approach towards the design of nonequilibrium HOWS and HONS phases, based on a time sequence of normal insulators. The obtained phases are unique in nonequilibrium systems because they cannot be described by bulk topological invariants used in time-independent situations.  We have instead advocated to characterize these phases with topological invariants under a dimension reduction framework.  The HOWS  and HONS phases discovered in this work still represent minimal complexity of each kind because they only contain two Weyl points or two nexus points, at either quasienergy zero or $\pi/T$.   As shown in  Supplementary Materials, coupled ring resonators in a 3D configuration \cite{Wang2016,Ochiai2016} should be a promising platform to experimentally confirm our main ideas.

\begin{acknowledgements}
J.G. acknowledge funding support by the Singapore Ministry of Education Academic Research Fund Tier-3 (Grant No. MOE2017-T3-1-001 and WBS No. R-144-000-425-592) and by the Singapore NRF Grant No. NRF-NRFI2017-04 (WBS No. R-144-000-378- 281).
\end{acknowledgements}

\clearpage
\onecolumngrid
\begin{center}
\textbf{\large Supplementary Materials}\end{center}
\setcounter{equation}{0}
\setcounter{figure}{0}
\setcounter{table}{0}
\newcommand{\cosinv}{\cos^{-1}}
\newcommand{\sininv}{\sin^{-1}}
\newcommand{\taninv}{\tan^{-1}}
\newcommand\mcF[1]{\multicolumn{4}{|c|}{#1}}
\newcommand\mcFF[1]{\multicolumn{2}{|c|}{#1}}
\newcommand\mcT[1]{\multicolumn{1}{|c|}{#1}}
\renewcommand{\theequation}{S\arabic{equation}}
\renewcommand{\thefigure}{S\arabic{figure}}
\renewcommand{\thetable}{S\arabic{table}}
\renewcommand{\cite}[1]{\citep{#1}}

\def\blue{\textcolor{blue}}
\def\red{\textcolor{red}}
\def\green{\textcolor{green}}
\def\bea{\begin{eqnarray}}
\def\eea{\end{eqnarray}}
\def\bal{\begin{aligned}}
\def\eal{\end{aligned}}

\newenvironment{rcases}
  {\left.\begin{aligned}}
  {\end{aligned}\right\rbrace}

\linespread{1.0}
\renewcommand{\theequation}{S\arabic{equation}}
\renewcommand{\thefigure}{S\arabic{figure}}

Supplementary Materials here are comprised of five small sections. In Section~I, we elaborate how to obtain the rich phase diagram plotted in Fig.~3a of the main text, starting from an explicit expression of the Floquet operator in the momentum space.  Section II discusses the chirality of the Weyl nodes for HOWSs in more details, followed by parallel discussions about the nodal lines for HONSs in Section III.  We then discuss the edge state spectrum and winding numbers of the sliced system in Section~IV. Finally, in Section~V we discuss a realistic platform of three-dimensional coupled ring resonators for future experimental confirmation of our main theoretical ideas.

\section{Phase Boundary of $2$D system}\label{App:Phase Diagram}
From Eq. (1, 2, 3) of the main text, we can write the Floquet operator as,
\bea\bal
U(k_{x},k_{y},k_{z}) = d_{0}\sigma_{0} - i [d_{x}\sigma_{x} + d_{y}\sigma_{y}  + d_{z}\sigma_{z} ]
~~~~\label{EQ:Floquet}
\eal\eea
where $\sigma_{x,y,z}$ are the Pauli matrices and $\sigma_{0} = \mathcal{I}_{2\times{2}}$. The coefficients of Pauli matrices are given as,
\bea\bal
d_{0} =& \cos^{3}(\gamma)\cos(\theta(k_{z})) + \sin^{3}(\gamma)\sin(\theta(k_{z})) - \cos(\gamma)\sin^{2}(\gamma)\cos(\theta(k_{z}))\left[\cos(k_{x}) + \cos(k_{y}) + \cos(k_{x} - k_{y})\right] \\
&- \cos^{2}(\gamma)\sin(\gamma)\sin(\theta(k_{z}))\left[\cos(k_{x}) + \cos(k_{y}) + \cos(k_{x} + k_{y})\right],\\
d_{x} =& \cos^{3}(\gamma)\sin(\theta(k_{z})) - \sin^{3}(\gamma)\cos(\theta(k_{z}))\cos(2k_{y}) - \cos(\gamma)\sin^{2}(\gamma)\sin(\theta(k_{z}))\left[\cos(k_{y}) + \cos(k_{x} + k_{y}) + \cos(k_{x} + 2k_{y})\right] \\
&+ \cos^{2}(\gamma)\sin(\gamma)\cos(\theta(k_{z}))\left[\cos(k_{x}) + \cos(k_{y})+ \cos(k_{x} + k_{y})\right],\\
d_{y} =& \sin^{3}(\gamma)\cos(\theta(k_{z}))\sin(2k_{y}) + \cos(\gamma)\sin^{2}(\gamma)\sin(\theta(k_{z}))\left[\sin(k_{y}) + \sin(k_{x} + k_{y}) + \sin(k_{x} + 2k_{y})\right] \\
& - \cos^{2}(\gamma)\sin(\gamma)\cos(\theta(k_{z}))\left[\sin(k_{x}) + \sin(k_{y}) + \sin(k_{x} + k_{y})\right],\\
d_{z} = & - \cos(\gamma)\sin^{2}(\gamma)\cos(\theta(k_{z}))\left[\sin(k_{x}) + \sin(k_{y}) + \sin(k_{x} - k_{y})\right] \\ &+ \cos^{2}(\gamma)\sin(\gamma)\sin(\theta(k_{z}))\left[\sin(k_{y}) - \sin(k_{x}) - \sin(k_{x} + k_{y})\right],
\label{EQ:Coefficients}
\eal\eea
The quasienergy of the system can then be obtained from Eq. (\ref{EQ:Floquet}) which is given as $\epsilon{T} = \cosinv[d_{0}]$. In order to determine the phase diagram as a function of $\theta(k_{z}) = \alpha + \beta\cos(k_{z}a)$ and $\gamma$, we first expand the Floquet operator around $\Gamma$ point $(k_{x_{0}}a, k_{y_{0}}a) = (0, 0)$  such that $(k_{x}a, k_{y}a) = ( k_{x_{0}}a + \delta_{x} , k_{y_{0}}a + \delta_{y})$. We consider first order approximation in $\delta_{x}$ and $\delta_{y}$ and the resulting Floquet operator is given as,
\bea\bal
U(\delta_{x}, \delta_{y}, \theta(k_{z})) =& \cos[3\gamma + \theta(k_{z})]\sigma_{0} - i\big[\sin(3\gamma + \theta(k_{z}))\sigma_{x} -\delta_{x}\sin(2\gamma)\sin(\gamma + \theta(k_{z}))\sigma_{z} \\
&- 2\sin(\gamma)[\delta_{x}\cos(\gamma)\cos(\gamma + \theta(k_{z})) + \delta_{y}\cos(2\gamma + \theta(k_{z}))]\sigma_{y} \big]
\label{EQ:FloquetT}
\eal\eea
The quasienergy $\epsilon{T} = \cosinv[d_{0}] = 3\gamma + \theta(k_{z})$ leads to the band crossing at zero quasienergy (red dotted line in Fig. 3(a) of main text) given as $3\gamma + \theta(k_{z})~ [{\rm modulo}, 2\pi] = 0$. Moreover, the condition of band closing at $\pi$ quasienergy (green dotted lines in Fig. 3(a) of main text) is given as $3\gamma + \theta(k_{z})~ [{\rm modulo}, 2\pi] = \pm{\pi}$. Similarly, the red solid in Fig. 3(a) of main text is given by the condition $\epsilon{T} = \cosinv[d_{0}] = \gamma - \theta(k_{z}) [{\rm modulo}, 2\pi] = 0$ and can be obtained by expanding the Floquet operator around $(k_{x_{0}}a) = \pi$ or $(k_{y_{0}}a) = \pi$ which are the nodal lines.

\section{Chirality of Weyl nodes}
From Eq. (\ref{EQ:FloquetT}), we can recognize that $k_{z}^{c} = \pm\cosinv[\frac{m\pi - 3\gamma - \alpha}{\beta}]$, and then the effective Weyl Hamiltonian $(H_{\rm eff} = -i{\rm log}[U])$ around $(0 , 0, \pm{k^{c}_{z}})$  is given by,
\bea\bal
H_{\rm eff} = m\pi\sigma_{0} \mp \sqrt{\beta^{2} - (m\pi - 3\gamma - \alpha)}\delta_{z}\sigma_{x} - [\delta_{x}\sin(2\gamma)\cos(2\gamma) + \delta_{y}\sin(2\gamma)]\sigma_{y} + \delta_{x}\sin^{2}(2\gamma)\sigma_{z},
\label{EQ:Weyl}
\eal\eea
where $m$ is a non-negative integer. Weyl nodes appear at zero [$\pi/T$] quasienergy for even [odd] values of $m$. The chirality of the Weyl nodes is then given as $\chi = \pm{\rm sign}[\sin^{3}(2\gamma)]$ \cite{Hosur2013} for $k^{c}_{z} = \pm\cosinv[\frac{m\pi - 3\gamma - \alpha}{\beta}]$ respectively. We can obtain Eq. (6) of the main text which is also given as $q = {\rm sign}[\frac{\partial{H_{\rm eff}}}{\partial{\delta_{z}}}] = \mp{1}$ from the effective Hamiltonian description and rightfully capture the charge of the Weyl nodes at $\pm{k^{c}_{z}}$. Moreover, it can be observed from Eq. (\ref{EQ:Weyl}) that the pseudo-spin components will be zero at $(k_{x}a, k_{y}a, k_{z}a) = (0, 0, \pm{k^{c}_{z}})$. We also notice that the $\sigma_x$ term is zero when $\delta_z=0$ in Eq. (\ref{EQ:Weyl}), which can be observed in Fig. 3(c) of the main text.

\section{Nodal lines}
From Fig. 3(a) of the main text and the discussion presented above, we have the condition of nodal lines given as $\gamma - \theta(k_{z}) [{\rm modulo}, 2\pi] = 0$ for $k_{x}a = \pi$ or $k_{y}a = \pi$. From this condition, we obtained that two nodal lines appear in the system for $k_{x}a = \pi$ or $k_{y}a = \pi$ at $k^{c}_{z} = \pm{\cosinv[\frac{\gamma - m\pi - \alpha}{\beta}]}$. In order to observe the quasienergy dispersion and the Berry phase based winding number around the nodal line, we expand the Floquet operator around $(k_{x}a, k_{y}a, k_{z}a) = (\pi + \delta_{x}, 0 + \delta_{y},\pm{k^{c}_{z}} + \delta_{z})$ and the effective Hamiltonian of the nodal line at $k_{y} = 0$ is given as,
\bea\bal
H_{\rm eff} = m\pi\sigma_{0} \mp \sqrt{\beta^{2} - (\gamma - m\pi - \alpha)^{2}}\delta_{z}\sigma_{x} + \delta_{x}\sin(2\gamma)\cos(2\gamma)\sigma_{y} + \delta_{x}\sin^{2}(2\gamma)\sigma_{z}
\eal\eea
where $m$ is a non-negative integer. Moreover, it can be observed that the effective Hamiltonian is independent of $\delta_{y}$ and linearly dependent on $\delta_{x}$ and $\delta_{z}$ around the nodal line. The winding number of ground state around the nodal line is then given from effective Hamiltonian as $W  = {\rm sign}[\frac{\partial{H_{\rm eff}}}{\partial{\delta_{z}}}]$. This will generate $W = \mp{1}$ at $\pm{k^{c}_{z}}$ for $\gamma = \alpha = 2\pi/3,~ \beta = \pi/2$.

\section{Winding numbers and topological edge states}

\begin{figure}
\includegraphics[width=0.8\linewidth]{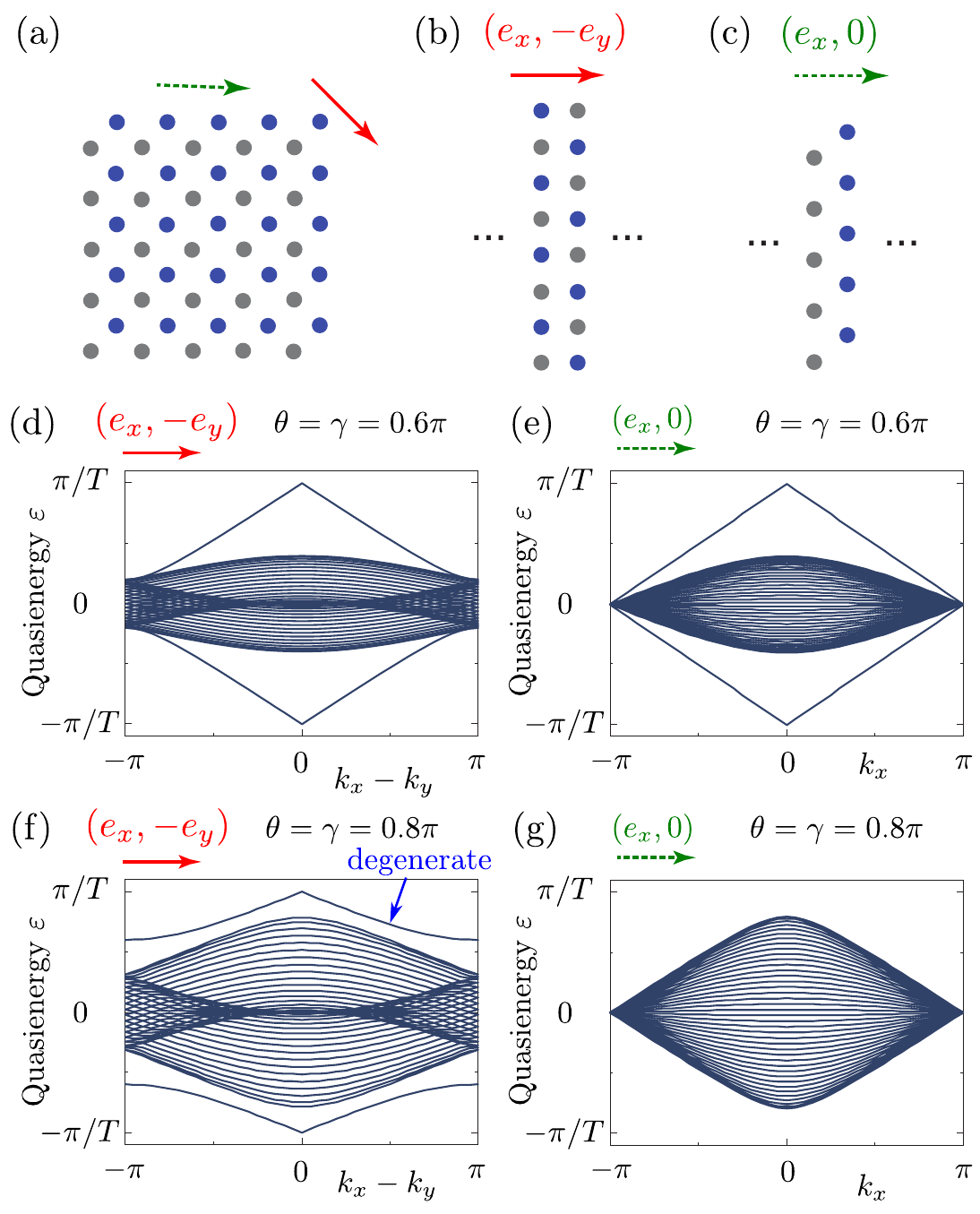}
\caption{Topological edge states of a reduced 2D system. (a) A finite structure of reduced 2D slices. (b) Semi-infinite structure with periodic boundary condition along $(e_x,-e_y)$ and open boundary condition along $(e_x,e_y)$. (c) Semi-infinite structure with periodic boundary condition along $(e_x,0)$ and open boundary condition along $(0,e_y)$. (d, e) Spectrum of semi-infinite structure with parameters $\theta=\gamma=0.6\pi$. (f, g) Spectrum of semi-infinite structure with parameters $\theta=\gamma=0.8\pi$. (d, f) have boundary condition as (b) and (e, g) have boundary condition as (c).}
\label{windingnumber}
\end{figure}

In this section, we use an example to show how the winding numbers determine the topological edge states and different topological phases. With fixed $k_z$, our system is reduced to a 2D system as shown in Fig.~\ref{windingnumber}(a), described by $\theta$ and $\gamma$. We study the spectrum of semi-infinite structure with different boundary conditions shown in Fig.~\ref{windingnumber}(b, c).  We first consider parameters $\theta=\gamma=0.6\pi$. In this case, the $\pi$ gap winding number is $1$. The spectrum of semi-finite structure with periodic boundary condition along $(e_x,-e_y)$ and open boundary condition along $(e_x,e_y)$ is shown in Fig.~\ref{windingnumber}(d). We notice that the spectrum contains chiral edge states. In particular, two $\pi$ modes are located at $k_x-k_y=0$ where the winding number is defined.  The winding number used to describe those topological states is that the system has chiral symmetry at $k_x-k_y=0$~\cite{Zhu2021z}.
 To check whether they are indeed chiral edge states, we consider another boundary condition which is periodic along $(e_x,0)$ and finite along $(0,e_y)$. The spectrum is shown in Fig.~\ref{windingnumber}(e).  There it is seen that chiral edge states still exist. We then consider another parameter $\theta=\gamma=0.8\pi$. In this case, the $\pi$ gap winding number is $2$. The spectrum of different boundary conditions are shown in Fig.~\ref{windingnumber}(f, g). For the first chosen boundary condition, the spectrum contains topological edge states as shown in Fig.~\ref{windingnumber}(f). Note that the obtained topological edge states are counter-propagating and degenerate. At $k_x-k_y=0$, there are four $\pi$ modes, consistent with the fact that the winding number we use to characterize the system is equal to $2$. Interestingly, now if we consider another different boundary condition, we find that the edge states are gapped as shown in Fig.~\ref{windingnumber}(g). These properties hint that the emergence of higher-order topological edge states (corner states in our case).

\section{More realistic model on an experimental platform}

\begin{figure}
\includegraphics[width=0.8\linewidth]{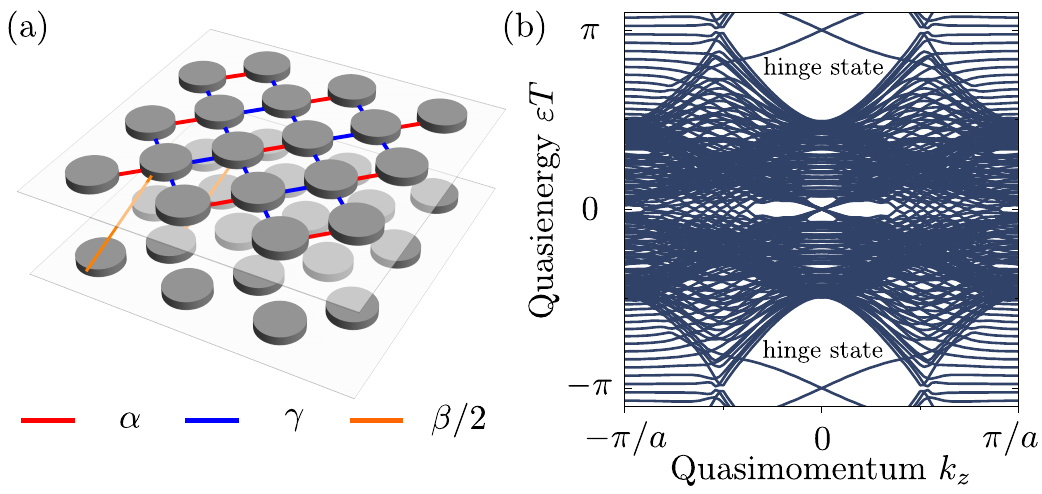}
\caption{Three dimensional coupled ring resonators and the spectrum obtained for a finite structure. (a) Schematic of coupled ring resonators. Red (Blue) line is the intra (inter) unit cell couplings. Orange line is the inter layer couplings. For simplicity, we only show the inter layer coupling in one unit cell. (b) Spectrum of a finite structure with periodic boundary condition along $z$ and fixed boundary condition along $x$ and $y$. The system contains 10 unit cells along $x$ and $y$. The parameters we choose are $\alpha=3\pi/4$, $\gamma=3\pi/4$ and $\beta=\pi/2$.}
\label{CRRs}
\end{figure}

In this last section, we discuss a more realistic model that is directly relevant to experiments. Indeed, 3D Floquet topological phases might sound difficult to realize, especially in cases with  low-frequency driving. In the main text, the model proposed is based on  coupled-layer construction.  Each layer has been proven to be equivalent to coupled ring resonators~\cite{Liang2013,Zhu2021z}. Indeed, such
two-dimensional coupled ring resonators have been used to experimentally measure topological invariant of anomalous Floquet topological insulators~\cite{Hu2015z}.  Extending this already available experimental platform along a third direction, we show in Fig.~\ref{CRRs}(a) an example of three dimensional coupled ring resonators.  This kind of configuration was previously utilized to study some Floquet Weyl semimetal phases~\cite{Wang2016z,Ochiai2016z}. Here we show that such a realistic platform can used to realize higher-order Weyl semimetal as we proposed.  Fig.~\ref{CRRs}(b) show spectrum of a finite structure with parameters $\alpha=3\pi/4$, $\gamma=3\pi/4$ and $\beta=\pi/2$. The method to obtain the spectrum can be found in Ref.~\cite{Wang2016z}, with the difference being that here we only dimerize the system.

Let us now focus on the $\pi/T$ quasienergy  gap.  It is noted that at $k_z=\pi/a$, the $\pi/T$ gap supports chiral edge state while at $k_z=0$, the $\pi/T$ gap supports topological corner states. At $k_z\approx \pi/(2a)$, the system features Weyl points. However, as a slight difference from the HOWS studied in the main text, here the hinge states are dispersive. That is, these hinge states can propagate along the $z$ direction. The reason is the following: the model system here based on a three-dimensional lattice of ring coupled resonators only has particle-hole symmetry and inversion symmetry, but not the two-fold rotation symmetry about $z$ axis, as shown in Fig.~\ref{CRRs}(a).  At fixed $k_z$, these symmetry properties combined still guarantee that the quasienergies of the topological corner modes are still located at $0$ or $\pi/T$.

\end{document}